\documentclass[sigconf]{acmart}
\AtBeginDocument{%
  \providecommand\BibTeX{{%
    \normalfont B\kern-0.5em{\scshape i\kern-0.25em b}\kern-0.8em\TeX}}}

\copyrightyear{2022} 
\acmYear{2022} 
\setcopyright{rightsretained} 
\acmConference[CHI '22 Extended Abstracts]{CHI Conference on Human Factors in Computing Systems Extended Abstracts}{April 29-May 5, 2022}{New Orleans, LA, USA}
\acmBooktitle{CHI Conference on Human Factors in Computing Systems Extended Abstracts (CHI '22 Extended Abstracts), April 29-May 5, 2022, New Orleans, LA, USA}
\acmDOI{10.1145/3491101.3519684}
\acmISBN{978-1-4503-9156-6/22/04}

\usepackage{caption}
\usepackage{subcaption}

\usepackage{amsmath}
\begin{document}

\title{Wizard of Errors: Introducing and Evaluating Machine Learning Errors in Wizard of Oz Studies}

\author{Anniek Jansen}
\email{a.jansen1@student.tue.nl}
\affiliation{%
 \institution{Department of Industrial Design, Eindhoven University of Technology}
 \city{Eindhoven}
 \country{The Netherlands}}

\author{Sara Colombo}
\email{s.colombo@tue.nl}
\affiliation{%
 \institution{Department of Industrial Design, Eindhoven University of Technology}
 \city{Eindhoven}
 \country{The Netherlands}}


\begin{abstract}
When designing Machine Learning (ML) enabled solutions, designers often need to simulate ML behavior through the Wizard of Oz (WoZ) approach to test the user experience before the ML model is available. Although reproducing ML errors is essential for having a good representation, they are rarely considered. We introduce Wizard of Errors (WoE), a tool for conducting WoZ studies on ML-enabled solutions that allows simulating ML errors during user experience assessment. We explored how this system can be used to simulate the behavior of a computer vision model. We tested WoE with design students to determine the importance of considering ML errors in design, the relevance of using descriptive error types instead of confusion matrix, and the suitability of manual error control in WoZ studies. Our work identifies several challenges, which prevent realistic error representation by designers in such studies. We discuss the implications of these findings for design.
\end{abstract}




\maketitle

\section{Introduction}
Machine Learning (ML) is becoming an increasingly important asset for designers as it provides, among others, new interaction possibilities (e.g. \cite{Laput2016Viband, Laput2017Synthetic}) and new personalization techniques (e.g. \cite{yang2016Planning, Yang2018_MapML, Gillies2016HCML}). ML provides the functional backbone of applications such as voice assistants, recommender systems, and face recognition. It can be embedded into consumer products in various ways, from narrow backend functionalities such as email spam filtering, to complex autonomous systems like robots or self-driving cars. 

Despite its potential, research has highlighted that designers find it challenging to use ML as a design material and to prototype with ML \cite{Yang2020, Yang2018_Investigating, Dove2017}, especially in the early phases of a design process. Training a custom ML model is a resource-intensive process and does not fit with the 'fail fast, fail often' approach typical of the design process early stages \cite{Dove2017}.
To overcome the difficulty of embedding working ML models in early prototypes to assess ML-enabled solutions, the Wizard of Oz (WoZ) method can be used \cite{Maulsby1993_prototypingwoz}. In this method, the wizard - a designer behind the screen who mimics the technology, can act like an ML model. The goal is to provide the user with the impression of interacting with a working system, and to test their reactions to the experience it generates \cite{Browne2019_WoZML}. 
This method is often used in the fields of design and HCI to prototype non-existing (or not yet realized) technologies and therefore represents a valid alternative to simulate the behavior of ML even when a model has not been trained yet. 
WoZ has already been applied to evaluate ML models in previous studies (e.g. \cite{Muller2019woz, Finke2019Lake, Viswanathan2020hybrid}). However, both these studies and additional research \cite{Allen2018_PrototypingAI, Browne2019_WoZML} show that simulating ML models behaviors in WoZ is no easy task. Among the challenges identified is the difficulty to realistically reproduce ML errors, because they are unlike human errors \cite{Begel2020_Lessons, Yang2019Sketching}. For instance, Riek \cite{Riek2012WOZHRI} shows that only 3.7\% of the WoZ studies in the Human-Robot Interaction domain include deliberate errors. ML errors are an intrinsic feature of ML models, and omitting them in a WoZ can lead to findings that are not representative of the user experience that is being simulated \cite{Begel2020_Lessons}.

\begin{figure*}
         \includegraphics[width=\textwidth]{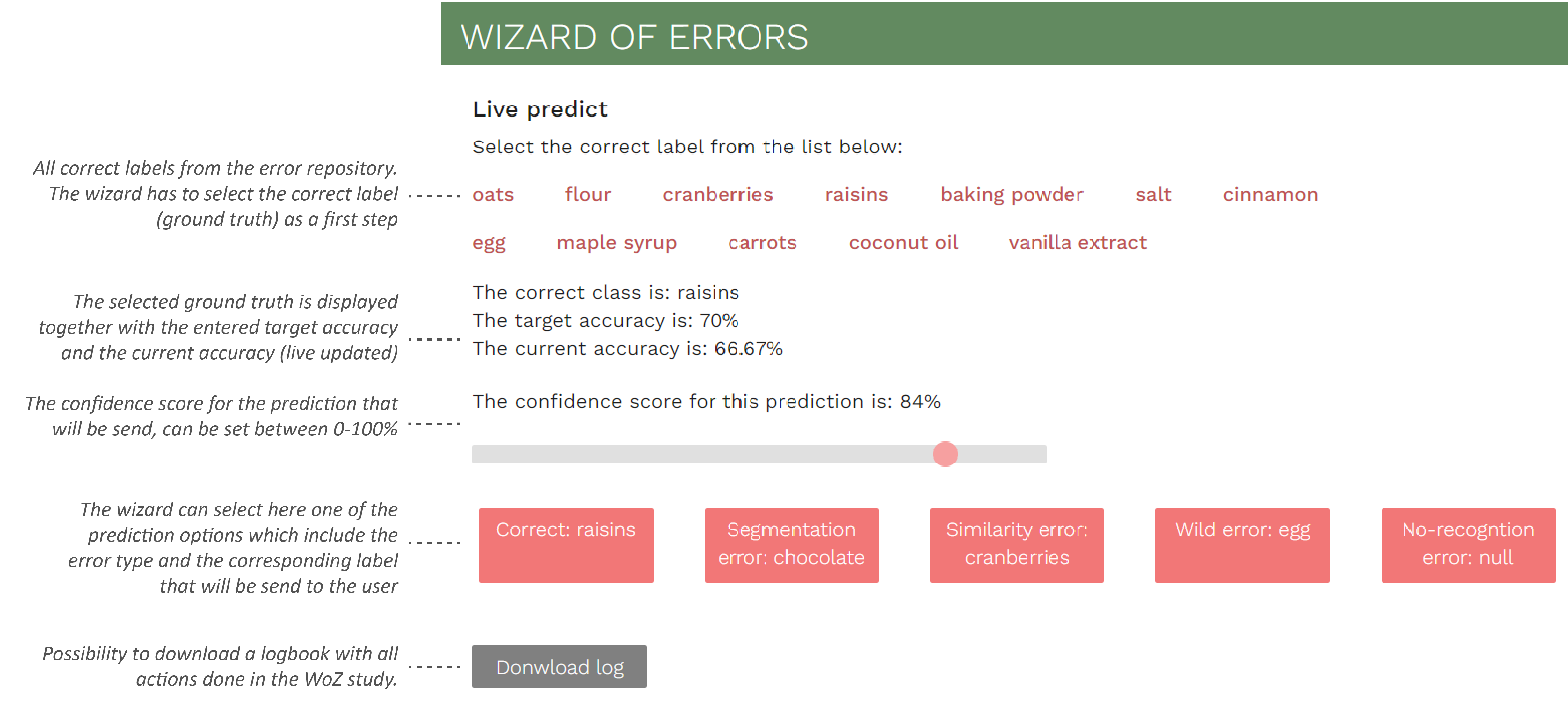}
         \caption{Prediction interface with explanations at the left-hand side}
         \label{fig: prediction interface}
         \Description{The live predict of the Wizard of Errors webpage. The page consists of a small header with the title. The elements on the site are annotated to explain there function. First there is a list of ingredients which are the ground truths with the annotation: All correct labels from the error repository. The wizard has to select the correct label (ground truth) as a first step. Next there is an overview of text witht the annotation: The selected ground truth is displayed together with the entered target accuracy and the current accuracy (live updated). Underneath there is a slider with the annotation: The confidence score for the prediction that will be send, can be set between 0-100\%. Then there are five buttons with different types of errors and labels inside of them with the annotation: The wizard can select here one of the prediction options which include the error type and the corresponding label that will be send to the user. Finally there is a last button with download log with annotation: Possibility to download a logbook with all actions done in the WoZ study.}
\end{figure*}

\paragraph{Advantages of considering ML errors early on}
Being able to successfully include ML errors in the early evaluation phases of a design process would allow designers to assess users’ reactions to different performance levels and errors of an ML model. This would have two main advantages. Firstly, it would allow designers to test the overall acceptability of a solution well in advance.  
Although the accuracy of an ML model is unknown before a model is trained, it can greatly influence the user experience (UX). For instance, in some applications like movie recommender systems, an 80\% accuracy score may be acceptable, while in others, such as in diagnostics systems based on image recognition (e.g. to detect skin conditions \cite{liu2020deep}) the same accuracy might be deemed unacceptable. Testing the impact of different accuracy levels on the UX early on would prevent the development of systems that are unsuccessful from a UX viewpoint.  

Secondly, testing different ML errors in advance would allow designers to create user interfaces that respond or adapt to ML errors, making the overall UX more pleasant or understandable. Previous studies have shown that ML errors can greatly impact the UX \cite{linguaFranca, Kocielnik2019ImperfectAI, Hong2021Planning}, the mental models users form to interact with the system, and their trust in the ML model \cite{Nourani2021Anchoring}. Different error types could differently influence the UX, e.g. categorizing a Golden Retriever as a Labrador might be perceived as an understandable ML error by users, while mistaking a dog for a bird might cause more frustration and distrust in the ML-based system. Considering different error types and testing their effects on the UX early on would inform design decisions to ensure that the system can 'fail gracefully' \cite{googlePAIR}, for instance giving users the possibility to report an error, or providing information on the types of errors that might occur, and their causes.

In this paper, we introduce Wizard of Errors (WoE), a web interface that facilitates the inclusion of ML errors in WoZ studies. This interface can be used in WoZ studies where the wizard can recognize the ground truth, i.e. the correct outcome of the ML prediction, from the input data. This happens, for instance, in object recognition, where the designer is able to recognize an object correctly, the same way an ML model is expected to do. 
We tested the WoE prototype with 10 design students to investigate if designers can successfully mirror ML errors during a WoZ study, in order to have a more representative simulation of the UX in the early phases of a design process.
This study aims to contribute to the current discussion on designing with ML, by showing how ML errors can be included and tested in WoZ studies, and by uncovering designers' challenges and misconceptions in mirroring an ML model behavior.

\section{Machine Learning failures and errors}
ML errors can occur in many forms and can have different origins. In this paper, we focus on errors that may occur during the user's interaction with ML-enabled solutions due to an incorrect ML prediction. During the remainder of this paper, the term \textit{ML error} will be used to describe such types of errors. 

A standard method to report ML errors is a confusion matrix. Here, errors are described as False Positives (FPs) and False Negatives (FNs). However, the confusion matrix, especially for a multi-class classification model (where the prediction is not a binary yes/no, but is based on three or more labels) \cite{tharwat2020classification} is hard to interpret for ML non-experts since the terminology is confusing and the reading direction and structure are unintuitive \cite{Shen2020Designing, Beauxis2018Supporting}. 
To overcome these issues and make errors more understandable to designers, we adopted the terminology introduced by Bott and Laviola \cite{Bott2015_WOZrecognizer} where they differentiate between four types of errors. The definition of the errors is tailored to their use case - classifying handwritten mathematical equations, but it can be generalized into the following definitions:
\begin{itemize}
    \item \emph{Segmentation error}: incorrect segmentation of the data that results in an incorrect prediction (e.g. recognizing a face in a photo in the background of a photo instead of the person in the foreground)
    \item \emph{Similarity error}: incorrect prediction that is somewhat related to the correct answer (e.g. recognizing sugar in a photo as salt)
    \item \emph{Wild errors}: incorrect prediction that appears to have no relationship with the correct answer (e.g. recognizing sugar in a photo as a carrot)
    \item \emph{No-recognition error}: failing to give a prediction (e.g. a face recognition system not responding when a face appears in front of the camera)
\end{itemize}

These descriptive error types can be used to create an error repository for an ML model, which contains all possible error options for each correct label. Such a repository can then be used in a WoZ study.
Compared to the confusion matrix, these four error types are expected to be easier to understand and to better support designers in testing the users' response to different types of errors.

\section{The Wizard of Errors Interface}
To enable the inclusion of the four ML error types in WoZ studies, we developed \emph{Wizard of Errors} (WoE). WoE is a web interface, which aims to help designers test the effects of different ML errors on users' experience while interacting with an ML-enabled system. The WoE application can be connected to different prototyping platforms (e.g. Processing, JavaScript, Python, Arduino) to perform WoZ studies. 
During the study, the WoE enables designers to simulate the predictions of supervised ML models by selecting either correct or incorrect labels (i.e. ML errors). For each correct label, four incorrect labels can be selected, based on the four different types of errors. Designers can also set an ML model target accuracy, so they can adjust the number of errors dynamically throughout the test, in order to reach the desired accuracy they intend to assess.

\begin{figure}

     \begin{subfigure}[t]{\columnwidth}
         \centering
         \includegraphics[width=\columnwidth]{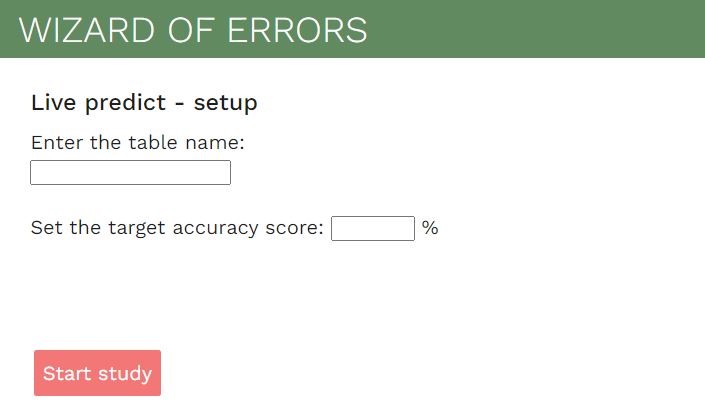}
         \caption{Setup page}
         \label{fig: setup}
         \Description{The setup page of the Wizard of Errors webpage. The page consists of a small header with the tile. Next a textbox where the table name of an uploaded error repository can be entered and a textbox for entering the target accuracy score. At the bottom of the page is a button to start the study.}
     \end{subfigure}
     \centering
     \begin{subfigure}[t]{\columnwidth}
         \centering
         \includegraphics[width=\columnwidth]{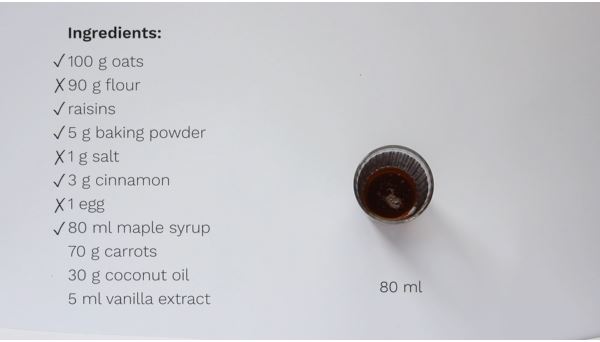}
         \caption{A screenshot of the video with ingredient list that was added to the prediction interface in Figure ~\ref{fig: prediction interface}}
         \label{fig: researchInterface}
         \Description{A screenshot of the video that was added to the prediction figure during the research. It shows a list of ingredients on the left side with a check mark, a crossmark or nothing in front of it and a glass with maple syrup with underneath 80 ml written on the screen. }
     \end{subfigure}
     \caption{Elements of the WoE interface}
        \label{fig: interface overview}
          
\end{figure}

The WoE application consists of two elements: (i) \emph{the error repository}; (ii) \emph{The WoE interface}. 
The error repository needs to be created by the designer in advance, as part of the WoZ setup, and it consists of a table that includes all the possible correct labels, and the corresponding incorrect labels - one for each of the four error types (see Table ~\ref{tab:errorRepository}). Once the error repository is uploaded to the WoE interface, the wizard-designer can decide to select a target accuracy, which will guide them in selecting the right amount of ML errors (see Figure ~\ref{fig: setup}).

The WoE interface is used during the WoZ study to simulate different errors during the user's interaction with the ML-enabled system (see Figure ~\ref{fig: prediction interface} for an overview of the system). 
The WoE interface is controlled by the wizard-designer, who can insert ML errors during the WoZ test, in order to determine how users react to potential ML errors. The interface allows designers to test three elements that potentially affect the overall UX: (i) the accuracy of the ML model; (ii) the types of ML errors that can occur; (iii) the confidence score for the predictions (a number between 0-1 or 0-100\% that indicates the probability associated to that label. A high confidence score does not mean that the label is correct, just that the machine associates that label to the input with a high probability).

\begin{table*}
  \caption{A selection from the error repository}
  \label{tab:errorRepository}
  \begin{tabular}{l|c|c|c|c|c}
    \toprule
     ID & correctAnswer  & segmentationError & similarityError & wildError & noRecognitionError\\
    \midrule
        0 & oats & cinnamon & flour & carrots & null\\
        1 & flour & salt & oats & maple syrup & null \\
    \bottomrule
\end{tabular}
\end{table*}

\section{Method}
To assess if designers could mirror the behavior of an ML model in a WoZ study, we tested the WoE interface with 10 design students. We aimed to investigate if designers would be able to simulate the ML behavior, including its errors, through the WoE interface, and what issues they would encounter. As a design case for our test, we used a smart kitchen countertop inspired by the IKEA concept kitchen  \cite{ikeaKitchen}, which adopts ML image recognition to identify ingredients placed on the countertop.
We asked participants to simulate the image recognition (ML) system performance through WoE, as if they were testing the user experience of the smart countertop in a WoZ study. Because in-person interactions were not possible, due to COVID-19 restrictions, the tests were performed online. In order to create the setting of a WoZ test, a video was generated, which simulated a sequence of user's actions, i.e. a series of ingredients being added to different bowls on the countertop. The video also showed an ingredient list on the screen. This video was shown next to the WoE interface (see Figure ~\ref{fig: researchInterface}) and participants were asked to use it as an input for simulating the ML system behavior in correctly or incorrectly classifying the ingredient being added to the bowl. We chose to simulate a user's interactions through a video because it allowed for consistency between the different tests - by providing all participants with the same inputs, and because in-person studies were not allowed.
For this study, an error repository was created by the researchers and uploaded to WoE. An extract of the error repository can be found in Table ~\ref{tab:errorRepository}.

\subsection{Participants}
Participants were selected through snowball sampling \cite{goodman1961snowball}. Students who were known to have worked with ML in a design project were approached, and they referred other students who also had experience with using ML in a design project. 10 participants were recruited (n=5 female; age: 22-27, M:23.5, SD: 1.71). All participants were Bachelor (n=1) or Master (n=9) Industrial Design students from a technical university. All of them had experience with using ML in at least one design project, up to six projects. 

\subsection{Procedure}
The study consisted of three parts and was conducted entirely online through a video-conference platform. The study protocol complied with the university Ethical Review Board procedures and all data were managed in accordance with GDPR regulations.
 
\paragraph{Part I - pre-experiment interview} For each subject, a pre-experiment semi-structured interview and a short survey were used to get insights into the participant's experience with and knowledge of ML errors.

\paragraph{Part II - the experiment} The design case of the smart countertop was introduced to the participant, together with the aim of the WoZ study, i.e. simulating the behavior of an ML system for image recognition, including its errors.
After describing the WoE interface, participants were asked to set the target accuracy score to 50\% and to start the video. The video showed an ingredient being poured into a bowl, and the task of the participant was to classify that ingredient, to simulate the ML prediction. After the ingredient was added, the video paused. The participant was then asked to (i) select the ground truth from the upper list (see Figure ~\ref{fig: prediction interface}), (ii) choose the confidence score for their prediction, and (iii) simulate the ML prediction by selecting either the correct label or one of the ML errors. If the correct label was chosen, a check mark would appear next to the ingredient in the video. In case of an error, a cross mark would be shown. This was meant to help the participant keep track of the number of correct predictions and errors they had simulated during the study. After recording the prediction, the video would continue by showing the next ingredient. The same procedure was repeated for 12 ingredients in total.

The WoE interface also showed the real-time accuracy score, based on the number of (in)correct predictions made up to that point. This allowed the participant to adjust their behavior over time, in order to reach the pre-defined target accuracy, i.e. make more correct predictions if the accuracy score was below the target accuracy score, or more incorrect predictions if the accuracy score was above the target.
Once the video ended, the participant was asked to follow the same procedure, watching the video again, this time with a target accuracy score set to 70\%.
Participants were asked to think aloud during the whole study, to explain the reasoning behind their choices and actions. All participants' interactions with WoE (i.e. logs) were recorded for subsequent analysis.

\paragraph{Part III - post-experiment interview} At the end of the experiment, a semi-structured interview was conducted and participants filled out a post-experiment survey. The interview and the survey were aimed to assess the participant's understanding of the error types, their view on how different error types could influence the UX, and what error they found most difficult to make.

\section{Preliminary results}
Results stem from a thematic analysis of qualitative data (i.e. audio recordings and interviews), as well as from the analysis of quantitative data (i.e. Likert scales from the survey and logs of the WoE system). Since the sample was small and homogeneous, we see these results as interesting initial findings that need to be validated through a larger study.

\subsection{Machine Learning errors}
Participants indicated that before working with ML, they were unaware of possible ML errors, but they became familiar with them during their first project. They assessed errors as increasingly important throughout the design process (Ideation phase: M=2.9, SD= 1.5; Realization phase: M=6.4, SD= 0.5). This assessment barely changed after the experiment (Ideation phase: M=3.0, SD= 1.7; Realization phase: M=6.5, SD= 0.5).

The descriptive types of ML errors used in this experiment were new to all participants. While most ML errors were clear, mainly due to their self-explanatory names, the segmentation error was more difficult to understand and only two participants could correctly define it afterwards. During the study, the participants could read the explanation by hovering over the buttons, but this option was hardly used.

Moreover, participants expected the different error types to have different impacts on the UX. While the similarity error was seen by most as less harmful because it was expected to be understandable by humans, the other errors were considered to have a more negative impact on the UX. Participants expected these errors to decrease the trust in the model or elicit frustration. \emph{"But if I put a carrot on the table and it is like “Oh strawberry” then I would be like, this is a stupid system it cannot even recognize such a simple thing then I must just work horribly"} (P2).

\subsection{Mirroring Machine Learning}
The main focus of the study was to evaluate if participants could mirror the behavior of an ML model by simulating ML errors. While all participants had worked with ML before and trained models themselves, their behavior did not match with that of an ML model. Several misconceptions surfaced during the experiment, as explained below.
Nevertheless, the WoE interface was successful in encouraging wizards to make ML errors and to reach the target accuracy score, although they were hesitant to make ML errors in the beginning (for $ accuracy=50(\%): M=59.10, SD=10.47;$ for $70(\%): M=68.72, SD=9.67$). Participants claimed that WoE was a useful tool to test the interactions in an early stage: \emph{"I think it would be helpful to explore the opportunities and limitations of ML in combination with a specific project. Normally this is difficult to do because in an early stage of a project a full system is often not working already."} (P6). However, they also indicated that it was hard to assess a realistic number of errors to include (P9) and that they would like to have something that could be applicable even earlier in the design process, to evaluate the potential impact of errors (P3, P4 and P5). 

\begin{figure}
     \begin{subfigure}{\columnwidth}
         \centering
         \includegraphics[width=\columnwidth]{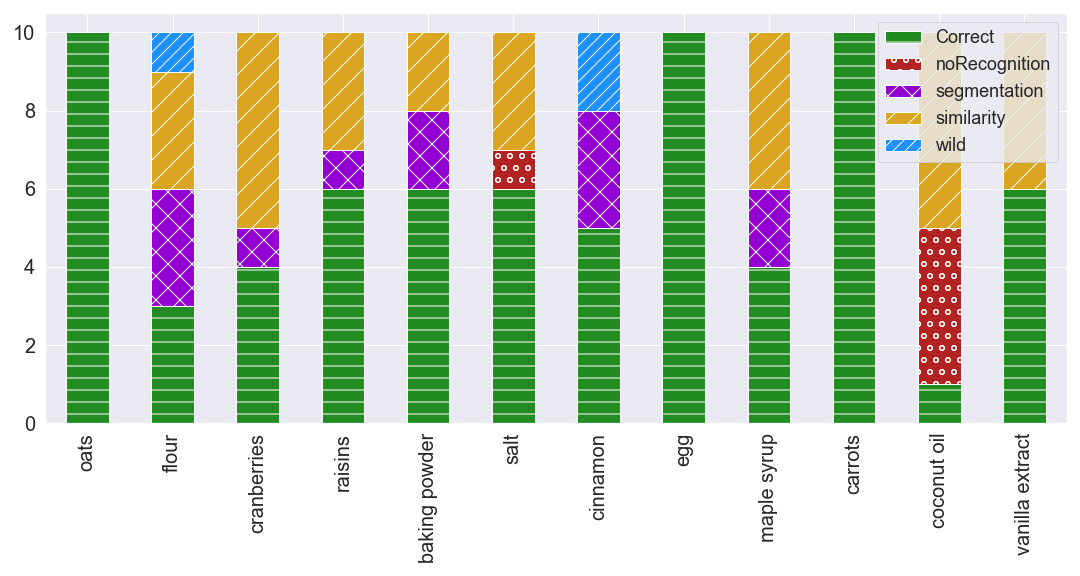}
         \caption{First cycle with target accuracy score 50\% }
         \label{fig:overview50}
         \Description{A stacked bar chart for the type of errors made for each ingrediens in case of a target accuracy score of 50\%. The majority of the predictions was correct, next a similarity error, a segmentation error, a no-recognition error and only three times a wild error. Both the oats, eggs, and carrots were predicted as correct by all participants. The coconut oil was only predicted correct by one participant.}
     \end{subfigure}
     \hfill
     \centering
     \begin{subfigure}{\columnwidth}
         \centering
         \includegraphics[width= \columnwidth]{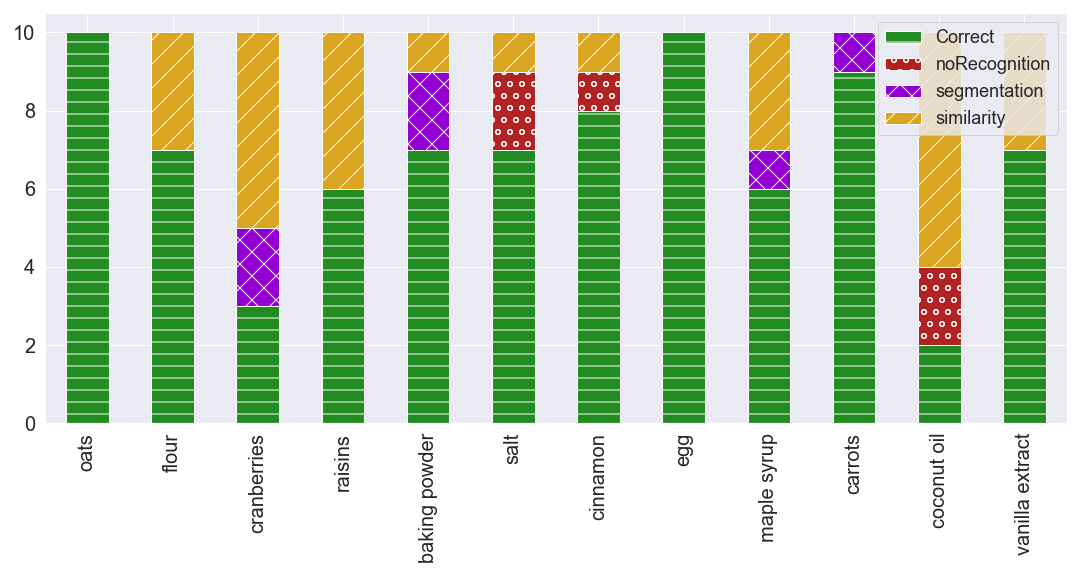}
         \caption{Second cycle with target accuracy score 70\%}
         \label{fig: overview70}
         \Description{A stacked bar chart for the type of errors made for each ingredients in case of a target accuracy score of 70\%. The majority of the predictions was correct, and more compared to the first cycle with target accuracy 50\%. The similarity error was made most often of the errors, next a segmentation error and finally a no-recognition error. There are no wild errors made. Both the oats and the eggs were predicted as correct by all participants. The coconut oil was only predicted correct by two participants.}
     \end{subfigure}
        \caption{Overview of the predictions for each ingredient}
        \label{fig: overview}
\end{figure}

\paragraph{Logical errors are easier to make}
The ML errors made during the experiment showed that participants mainly selected similarity errors and hardly any wild errors (see Figure ~\ref{fig: overview}). From the think-aloud transcripts, it became clear that participants based their choices mainly on the label associated to each error type and not the error type itself (e.g. choosing the 'sugar' label for salt, instead of choosing a similarity error). By using labels, participants often selected an understandable or logical error from a human perspective. They also expressed it felt unnatural to send an unrelated error.
As participant 9 stated:
\emph{"For me as a wizard, a no-recognition error would be weird to make because I know what it is and rationalizing why it would not recognize it at all would be a bit weird, in my opinion."}

\paragraph{Projecting personal knowledge}
Similarly to the tendency to select logical errors, participants based the behavior of the ML model on their own knowledge. This was mainly visible when coconut oil was shown in the video, and 85\% of the predictions were an ML error - while the mean for all other ingredients, excluding coconut oil, was 31.8\%. As participant 7 put it: \emph{"I myself have never used coconut oil before, so I am not really sure if it is an oil or if it looks like this."}

The tendency to rely on their own knowledge could also be seen in participants' assumption that common ingredients (in their food culture) are easier to recognize: \emph{"If there are ingredients that are exotic, then it might be easier to make [a no-recogntion error]"} (P1). Moreover, they did not consider that ingredients can exist in different shapes and forms. Eggs and carrots were seen as very easy to classify for the machine because they had a distinctive shape, but participants did not take into account that they could exist in many forms and shapes, which would make it harder for the ML model to recognize them correctly. 

\paragraph{Higher confidence score for correct answers}
To check if participants gave higher confidence scores to correct predictions, a linear regression was calculated. A significant regression was found ($F(1,238)=35.618, p<.001$), with an ${R}^2$ of $.130$ and an ${R}^2_{Adjusted}$ of $.127$, indicating that the error type (correct vs incorrect) is a significant predictor for the confidence score ($\beta= .361, t(238) = 5.968, p < .001$) and that the confidence score increased when a correct prediction was selected. This differs from a real ML model, where a confidence score can be equally high for incorrect predictions.

\paragraph{ML model self-awareness}
Finally, participants expected the model to be aware that it was making a mistake and mentioned that it would be useful if it would indicate to the final user what type of error it was making: \emph{"I think for the user it is very helpful to know that there is an error, and they know what kind of error it is."} (P3). 
A possible explanation for this misconception might be that the participants had only trained and tested an ML model and not deployed it live, since in training and testing supervised classification models the correct label is available and makes it possible to detect errors - although not the error types.
\section{Discussion}
The preliminary results presented in the previous section provide insights into what aspects of mirroring an ML model are difficult for wizard-designers.
These findings have implications both for designing WoZ studies for ML applications, and for designing with ML in general. 

When designing WoZ studies to test user experiences and interactions based on ML, one needs to take into account the fact that wizards will not be able to ignore their human rationale, therefore their behavior will not be a good representation of the ML behavior. Our study showed that the WoE interface and the ML errors can already contribute to a better representation of ML models in WoZ studies and can trigger designers to reflect on the importance of testing different types of ML errors. However, it also is clear that this is not sufficient, as designers' misconceptions prevent them from properly mirroring ML models. 
One way to overcome this limitation is to recommend wizards certain error types during the use of the WoE interface, for instance in case certain error types are more rarely selected, compared to others. 

Another option is to allow designers to pre-define how many errors for each type will be sent by the WoE interface, and let the interface randomly assign those errors, while asking designers only to select the ground truth. While limiting designers' ability to adjust the number and type of errors in real-time, this would make the study more realistic.

Next to that, designers not only need to test different error types, but also vary on the number and the moment of occurrence to explore the effect of all these factors on the user's experience and their trust in the ML predictions. 
While this study only focused on using WoE in the early phase of a design process, ML errors can also be introduced during other phases. For instance, in the ideation and conceptualization phase, cards with the error types, potential consequences and options for adjusting the design can help to consider errors from the beginning. When designing a UX wireframe, the designer can go over each step and consider what would happen if each type of error occurred, to check if their interface is able to fail gracefully \cite{googlePAIR} or if it needs to be adjusted so that it can fail gracefully. 

\section{Limitations and Future Work}
A limitation of the WoE interface, and WoZ testing with ML in general, is that it is only applicable in cases where the humans are able to make the  correct predictions - i.e. recognize the ground truth, during the study. However, this still leaves out a considerable number of ML models within supervised learning.
This study  only covered object recognition, but we expect the error types to be also transferable to other subsections of computer vision and potentially also to other types of supervised classification. Further research should explore and validate the use of ML in these applications. 

Another limitation of this study is that it was conducted with students from one Industrial Design faculty, which could give a biased view on how design students, and designers in general, approach and use ML. Therefore, we consider these findings as preliminary results and future studies should be conducted with a larger and more diverse sample to confirm their validity. Furthermore, the WoE interface should be tested in a WoZ study with actual users, to gather more insights on designers' behaviors in a real context. 

\section{Conclusion}
Using ML in prototypes during the early phases of a design process is challenging. In this study, we introduced Wizard of Errors (WoE), a prototyping tool for conducting WoZ studies on ML-enabled interactions. The WoE interface  facilitates the inclusion of ML errors into WoZ studies, since these are essential for UX assessment but are currently rarely included. In WoE we used four descriptive error types instead of the confusion matrix and evaluated with design students if ML errors are important to consider and how relevant the four error types are in comparison to the confusion matrix. Moreover, during a WoZ simulation, we evaluated if it is possible for a designer to simulate an ML model in terms of ML error behavior. 
The error types showed to have potential to be used during WoZ studies, but we also identified several challenges that still prevent the designer from realistic error representation in WoZ.
In this study, we designed and tested the WoE interface to embed ML errors in WoZ studies, and provide preliminary knowledge that can help both design researchers and practitioners in this field to consider ML errors as a regular component of the design process for ML-enabled solutions.


\bibliographystyle{ACM-Reference-Format}
\bibliography{sample-base}


\end{document}